# An Interpretable Operator-Learning Model for Electric Field Profile Reconstruction in Discharges Based on the EFISH Method

**Zhijian Yang[1*], Edwin Setiadi Sugeng[1], Mhedine Alicherif[2] and Tat Loon Chng[1]**

[1] Department of Mechanical Engineering, College of Design and Engineering, National University of Singapore, 117575, Singapore
[2] Mechanical Engineering Program, PSE, King Abdullah University of Science and Technology (KAUST), Thuwal 23955-6900, Saudi Arabia

*Author to whom any correspondence should be addressed.

E-mail: zhijiany@nus.edu.sg

## Abstract

Machine learning (ML) models have recently been utilized for reconstructing the electric field distribution from a corresponding EFISH signal profile – a task described in earlier works as the 'inverse EFISH problem'. This approach is one of several that addresses the inaccuracy of a (line-of-sight) EFISH measurement caused by the Gouy phase shift present in focused laser beams. A key advantage of this approach is that the accuracy of the reconstructed (or 'inverted') profile can be readily evaluated through a 'forward transform' of the underlying EFISH equation. Motivated by this latest success, the present study introduces a novel ML model with significantly enhanced performance. This model is built on a more complex, but powerful operator-learning architecture, and moves beyond the simpler class of artificial and convolutional neural networks (i.e., ANNs and CNNs) employed in previous work. Termed here as Decoder-DeepONet (or DDON), its main asset is the ability to learn function-to-function mappings, a feature essential for recovering electric field profiles of an unknown shape. The superior performance of DDON is exemplified via a comparison with our published CNN model and the feasibility of a classical mathematical method, as well as its application to both discharge simulations and experimental EFISH data from a nanosecond pulsed discharge. In almost all cases, the DDON model exhibits better generalizability, higher prediction accuracy, and wider applicability. Furthermore, the intrinsic nature of this operator-learning architecture renders it less sensitive to the exact location(s) of the acquired data, enabling electric field reconstruction even with seemingly 'incomplete' input profiles – an issue often accompanying poor signal sensitivity. Another important aspect of this work is the use of 'Integrated Gradients' (or IG), which helps identify input (signal) regions that most critically influence the accuracy of a reconstructed profile. This quantitative metric can in turn provide guidance on the optimal sampling region (or window) for acquiring EFISH data. Overall, we believe that the DDON model is a robust and comprehensive model which can be readily applied to reconstruct 'bell-shaped' electric field profiles with an existing axis of symmetry, especially in non-equilibrium plasmas.

## 1. Introduction

Measuring electric fields in discharge plasmas is essential, as the field initiates, and thus governs, important processes such as charge and species generation, recombination and transport—which ultimately shape plasma chemistry and dynamics. As such, direct field measurements remain indispensable for validation and capturing experimental realities such as electrode geometry, transient phenomena, and compositional impurities. Conventional field probes, however, perturb the local electric field when introduced into the discharge, distorting the very quantity to be measured. This has driven sustained interest in non-contact, non-intrusive diagnostics capable of resolving rapid spatiotemporal field variations without altering the plasma environment.

Electric-field-induced second-harmonic generation (EFISH) has emerged as one such promising solution. First observed in the mid-1970s by Bigio et al. [1] as a nonlinear optical effect in centrosymmetric medium, EFISH has recently been developed into a species-independent, non-resonant, and non-contact diagnostic with excellent temporal resolution for electric field measurements in non-equilibrium plasmas [2,3]. Conceptually, a focused laser probe of power $P^{(\omega)}$ propagates through a plasma along the line-of-sight of the beam path (viz. z-axis). The nonlinear interaction between the applied electric field $E_{\mathrm{ext}}(z)$ (of this plasma) and the optical electric field (of the laser beam) in turn generates a coherent second-harmonic signal $P^{(2\omega)}$. The EFISH signal power $P^{(2\omega)}$ scales quadratically with the applied electric field through an integral that incorporates both properties of the plasma as well as phase-matching and focusing effects of the probe laser [4,5] as shown below:

$$P^{(2\omega)} \propto [\alpha^{(3)} \cdot N \cdot P_{\mathrm{o}}^{(\omega)}]^2 \cdot \frac{1}{z_{\mathrm{R}}} \cdot \left| \int_{-\infty}^{\infty} \frac{E_{\mathrm{ext}}(z) \cdot e^{i \cdot \Delta k \cdot z}}{1 + i \cdot \frac{z}{z_{\mathrm{R}}}} \, \mathrm{d}z \right|^2, \tag{1}$$

where $\alpha^{(3)}$ is the third-order nonlinear hyperpolarizability (a fourth rank tensor), $N$ is the gas number density, $P^{(\omega)}$ is the probe beam power, $P^{(2\omega)}$ is EFISH signal power, $z$ is the beam propagation axis, $z_{\mathrm{R}}$ is the laser beam Rayleigh range, $E_{\mathrm{ext}}$ is the externally applied electric field, and $\Delta k$ is the wave-vector (phase) mismatch. (Both the beam focus and the origin of $E_{\mathrm{ext}}$ are located at $z = 0$.)

A fundamental implication follows directly from the above axial ($z$) integral: since the second harmonic builds coherently along the (beam) propagation direction, the resulting EFISH signal is sensitive to both the Gouy phase and phase mismatch, and depends on the entire axial shape of $E_{\mathrm{ext}}(z)$ rather than exclusively on its local value within the focal region. An unknown field-profile shape thus introduces additional measurement uncertainty—precisely because the signal integrates along $z$ beyond the coherence length with a nonlinear weighting. These factors, discussed in detail elsewhere [4,6], render naive quantification of $E_{\mathrm{ext}}$ from $P^{(2\omega)}$ inaccurate and instead require further treatment to address this issue. Recent research has identified two potential solutions to this problem, each with their respective benefits and limitations. The first involves the use of crossed laser beams, intersecting at a fixed angle, to isolate the interaction region, thereby avoiding the line-of-sight buildup error [7–10]. As discussed more extensively in Ref. [11], in this approach lies a fundamental trade-off between spatial resolution and detection sensitvity, one which has prompted further research into various methods of signal amplification [12]. The second approach, which is adopted in this work, relies on acquiring an EFISH signal profile, $P^{(2\omega)}(z)$, by translating the plasma with respect to the beam focus. This signal profile can in turn be used to reconstruct the shape of the unknown electric field profile, $E_{\mathrm{ext}}(z)$**.** Given that the process of recovering of $E_{\mathrm{ext}}(z)$ from $P^{(2\omega)}(z)$ entails, in our opinion, a mathematically non-trivial problem exacerbated by a loss of phase information, this has motivated research into this underlying 'inverse EFISH problem'.

Several strategies have been proposed for EFISH inversion. One line of work exploits axisymmetry to recast the problem as an inverse-Abel transform requiring deconvolution [13,14]. Other approaches target inhomogeneous fields via machine learning and numerical optimization (e.g. SQP) [15], but are largely limited by its repeatability and sensitivity to initial profiles and tuning choices, while similar limitations arise in discrete inversion schemes [16]. A more recent publication has focused on recovery of phase information to perform this reconstruction [17].

A complementary, data-driven route is machine learning, which dispenses with strong priors and rule-based fitting, and is both flexible and computationally efficient [11]. In earlier work, we introduced a convolutional neural network (CNN) that predicts $E_{\mathrm{ext}}(z)$ directly from the measured EFISH profile $P^{(2\omega)}(z)$ [11]. The approach achieved excellent accuracy on synthetic datasets and experimentally acquired electrostatic fields—even with sparse sampling (see Ref. [11]). However, for various reasons detailed below, its generalizability – defined broadly as the accuracy of an ML model in making predictions on new, unseen (input) data that is not part of the training set – could be limited in practice.

- **Lack of prior shape knowledge.** In many settings, one typically does not know a priori which function family best represents the true field profile, $E_{\mathrm{ext}}(z)$. The CNN model in Ref. [11] was trained solely on profiles which are part of the Fuzzy family (i.e., equation (4) of Ref. [11] or see Table 2). This was substantiated by empirical evidence that these profiles accurately match the *electrostatic* field distribution produced by canonical geometries. Nonetheless, constraining the model to a particular profile type risks systematic bias if the actual field shape lies outside this training family. We add that most published work adopting this reconstruction approach have either been limited to electrostatic fields, or profiles (of $E_{\text{ext}}(z)$ or $P^{(2\omega)}(z)$ ) corresponding to a few selected shapes: such as Lorentzian or Gaussian [18,19]. Consequently, none of these studies have surveyed the broader function space, which *may* very well occur in a plasma.

- **Profiles across different function families.** In line with the above, model performance degrades when tested $E_{\text{ext}}(z)$ profiles depart significantly from the Fuzzy family used for training. Examples of such profiles include exponential or Voigt-like shapes shown in Appendix B. That actual profiles could possibly be different is plausible: electric-field morphologies reflect intertwined influences—electrode geometry, dielectric boundaries, gas composition and pressure, temperature, streamer/ionization dynamics, optical parameters and phase matching, and measurement noise—each capable of modifying or distorting the profile shape [18,19].
- **Absence of clear acquisition guidelines.** Experimentally, there is no universal criterion for the spatial window/region that must be sampled to achieve robust and accurate inversion of $P^{(2\omega)}(z)$ to $E_{\text{ext}}(z)$. There is thus a lack of *explicit* guidance on the measurement extent (especially relative to optical parameters such as the Rayleigh range) required to recover $E_{\text{ext}}(z)$ with bounded error.

Taken together, the above factors motivate a more flexible operator-learning formulation that (a) is explicitly robust across multiple function classes/families of $E_{\text{ext}}(z)$, and (b) provides clear acquisition guidelines—such as sampling extent and density—to enable reliable profile inversion ($P^{(2\omega)} \rightarrow E_{\text{ext}}$) under realistic experimental conditions.

Recently, the Deep Operator Network (DeepONet) has gained considerable traction for learning mappings between function spaces, owing to its improved architectural flexibility and generalizability compared to a regular fully-connected neural network [20,21]. Its branch-trunk design is naturally built for accommodating heterogeneous inputs (e.g., profile shapes and other optical parameters). In the present context of EFISH inversion, this permits the separation of profile information, $P^{(2\omega)}$, from other parameters such as the location of acquired data, $z$, as well as optical parameters such as $\Delta k$. That is, this intricate architecture decomposes the EFISH integration function into two independent spaces (or latents) and parameterizes them through a bunch of networks; the branch latent learns a compact set of feature coefficients for the $P^{(2\omega)}$, while the trunk learns a corresponding set of bases conditioned on the integration and thus $E_{\text{ext}}(z)$. However, their connection is constrained to a simple dot product—a fixed linear handshake rather than a learned nonlinear transformation. This rigidity compels each network to develop a shared and *disentangled* representation where the branch captures essential features of the profile and the trunk learns to modulate them in response to the electric field. This not only avoids 'interference' between these different parameters but provides better input flexibility and response accuracy to these (multiple) interdependent variables. In practice, DeepONet has been successfully deployed in a range of complex systems, including stress distribution prediction [22], flow-field inference around airfoils [23], and plasma sheath dynamics in semiconductor processing [24], among others. Building on these advances, we pursue an optimized operator-learning formulation (Decoder-DeepONet or DDON) tailored to the problem of EFISH inversion, with two complementary objectives: cross-family and noise robustness, and practically useful guidance for data acquisition.

Furthermore, with the immense popularity of machine learning studies in physics, understanding how trained models make decisions has become a central concern. Rather than treating models as black boxes, explainable-AI (XAI) techniques are increasingly used to build trust and inject physical insight [25]. Therefore, we complement our model's predictive capability with the use of Integrated Gradients (IG) [25,26], a popular method that produces saliency maps highlighting which portions of the input signal most strongly drive the predictions – a process known as 'attribution'. In our study, IG serves two practical roles: it provides insight into the model's decision pathways with respect to different profile locations and shapes (e.g., peak vs. tails; single- vs. double-peak profiles) and furnishes actionable guidance for data acquisition by quantifying a conservative, normalized "key" sampling window around the laser focus. The XAI-informed window streamlines experimental planning and helps explain DDON's decision-making process, tying operator learning to physically meaningful measurement cues.

The sections that follow describe the specifics of the new DDON model and the working principles of IG, an evaluation of the model's performance when compared against our previous CNN model, (ii) its noise robustness, and (iii) its prediction accuracy when applied to existing simulation and experimental datasets.

## 2. Methodology

### *2.1 Deep Operator Network (DeepONet)*

DeepONet [20] is a neural architecture expressly designed to learn operators between function spaces, often outperforming pointwise regression approaches (e.g., conventional deep neural networks, or ANNs) that target fixed input–output mappings. The standard formulation comprises two subnetworks: (i) a branch network that encodes the discretized input function into a latent vector $\beta(x)$ evaluated at points $x = \{x_1, x_2, \dots, x_m\}$; and (ii) a trunk network that encodes the observation location(s) $y = \{y_1, y_2, \dots, y_n\}$ of the output or input function into a latent vector $\tau(y)$. The two streams are combined by a dot product—yielding the operator's action at the queried location(s). In essence, DeepONet learns a mapping that takes an input function (through $\beta$) and returns an output function evaluated at $y$ (through $\tau$) [21].

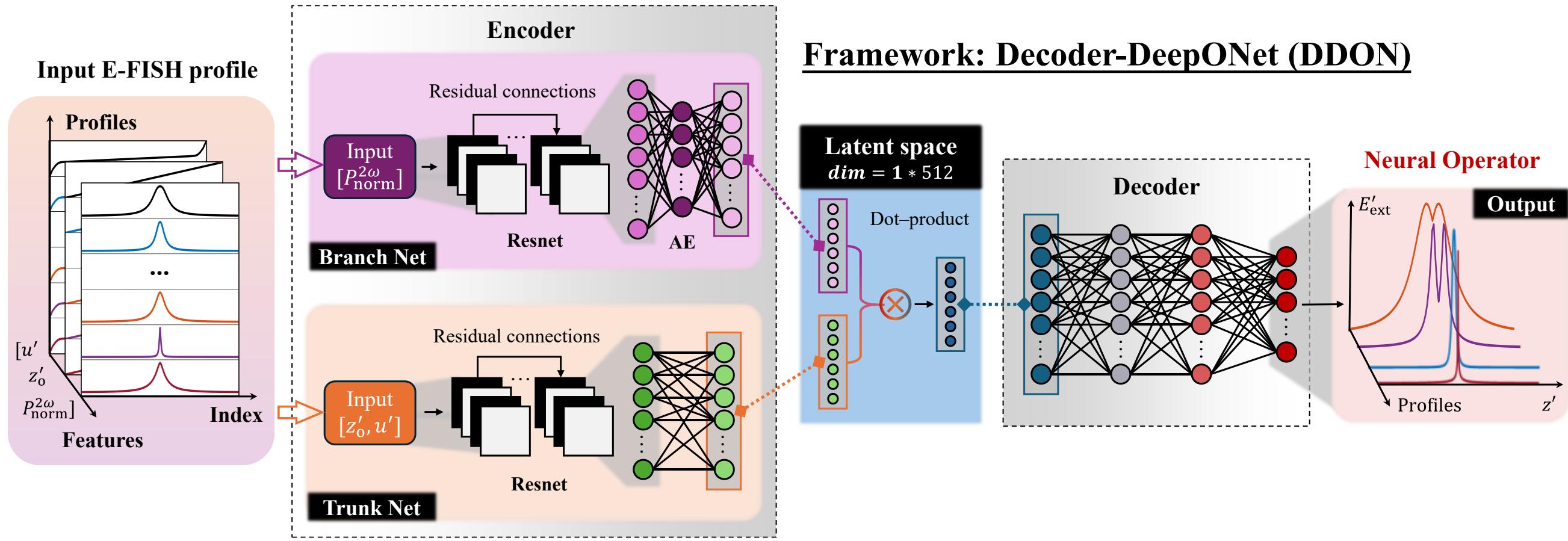

Figure 1 Decoder-DeepONet (DDON) architecture for operator learning: inputs ($P^{2\omega}_{\text{norm}}$ for branch net and $[u', z'_0]$ for trunk net) and outputs (electric field, $E'_{\text{ext}}(z')$) are mapped through an encoder that couples a ResNet with an autoencoder (AE, for branch net). The branch net uses three residual blocks with filters $\mathcal{F} = \{64, 128, 256\}$; the AE compresses to a 32-dimensional bottleneck (capturing global context and key local variations) and yields a 512-dimensional branch latent. The trunk net mirrors the branch-side ResNet and is followed by dense layers (width {512}) to produce a 512-dimensional trunk latent. The decoder fuses branch and trunk latents via fully connected layers to generate output profiles. $P^{2\omega}_{\text{norm}}$ denotes the normalized EFISH signal, while $[u', z'_0]$ denote non-dimensionalized wave-vector mismatch and expected measurement locations of electric field, respectively. Note: the range and values of $z'$ for describing $E'_{\text{ext}}(z')$ are identical to $z'_0$, $z' \equiv z'_0$.

Formally, the output of the learned operator can be written as:

$$\mathcal{G}(\beta)(\tau) \approx \sum_{m=1}^{d} \beta_m(x)\, \tau_m(y), \tag{2}$$

where $\{\beta_m\}_{m=1}^{d}$ and $\{\tau_m\}_{m=1}^{d}$ denote the latent features produced by the branch and trunk networks, respectively, and $d$ is the latent dimension.

## *2.2 Decoder-DeepONet (DDON)*

DeepONet [21] is a flexible operator-learning framework whose branch and trunk subnetworks can adopt diverse neural architectures. Building on this versatility, we tailor DeepONet to EFISH inversion and further introduce an encoder-decoder design [23,27]. Inherited from DeepONet, the preserved branch-trunk architecture in DDON allows the EFISH–to–E-field mapping ($P^{(2\omega)} \rightarrow E_{\text{ext}}$) and the (optical/location) parameters to be represented in distinct latents—learned separately yet optimized jointly—thereby reducing interference. For EFISH inversion tasks where the input is a distribution (function) rather than a vector, the learned operator provides a more faithful, data-efficient approximator than standard neural networks. The encoders are introduced to further stabilize the mapping, fuse multi-scale features, and improve generalization across function families. In brief, our DDON compresses salient patterns into a compact latent via the encoder and reconstructs an operator-aware representation via the decoder.

While CNNs excel at extracting local and global features [11], deeper stacks can suffer vanishing gradients and degradation. We therefore employ ResNet encoders [28] alongside a vanilla autoencoder (AE) to process EFISH signals and E-field observation/parameter streams. The dataset spans canonical bell-shaped profiles (Voigt, Fuzzy) with realistic variations [6], providing a controlled yet diverse training ground. In this setting, ResNet depth enables richer, more discriminative representations, yielding stable and accurate inversion; empirically, the ResNet-AE pairing offers a practical balance—capturing cross-family variability while limiting overfitting and training pathologies.

As depicted in Figure 1, the model takes $\{[P^{2\omega}_{\text{norm}}(z'_0),\ z'_0],\ [u']\}$ as inputs and predicts the electric field profile $E'_{\text{ext}}(z')$ (data details are discussed in § 2.3). The branch network processes the EFISH signal with a ResNet + AE stack. The ResNet comprises three residual blocks with filter counts $\mathcal{F} = \{64, 128, 256\}$; its final feature map is passed through global average pooling and flattened, then fed to a vanilla AE (layer widths {1024, 128, 32, 128, 512}). The AE compresses the representation to a 32-dimensional bottleneck (capturing global context and salient local variations) and produces a 512-dimensional branch latent $\beta(P^{2\omega}_{\text{norm}}) \in \mathbb{R}^{512}$, which serves as the operator-facing embedding for the branch stream. In parallel, the trunk network encodes the observation locations and configuration parameters $\{z'_o, u'\}$ using the same ResNet backbone (initial 1D convolution, followed by three residual blocks with strided downsampling), global average pooling, flattening, and a projection through three dense layers (widths $\{512\}$) to yield a 512-dimensional trunk latent $\tau(z'_o, u') \in \mathbb{R}^{512}$. Throughout, "latent" refers to these 512-dimensional embeddings produced by the branch and trunk encoders, which are fused by the operator head $\mathcal{G}$.

The operator head computes a dot-product coupling between the branch and trunk latents:

$$\mathcal{G}(\beta,\tau) = \left\langle \underbrace{\beta(P^{2\omega}_{\text{norm}})}_{\text{branch}}, \underbrace{\tau(z'_o,u')}_{\text{trunk}} \right\rangle = \sum_{m=1}^{512} \beta_m \tau_m, \quad (3)$$

yielding a 1D latent response. A decoder then "unembeds" this 1D vector back to the profile coordinates via a sequence of dense layers with widths $\{512, 512, 256, 256, 109\}$, producing the final prediction. For the dense layers, we apply dropout (rate 10%) and L2 regularization ($1 \times 10^{-7}$) to curb overfitting and enhance generalization [29–32]. The overall operation can thus be written as

$$\mathcal{D}\big(\mathcal{G}(\beta,\tau)\big) = \underbrace{\mathcal{D}}_{\text{decoder}}\left(\left\langle \underbrace{\beta(P^{2\omega}_{\text{norm}})}_{\text{branch}}, \underbrace{\tau(z'_o,u')}_{\text{trunk}} \right\rangle\right) = \underbrace{\mathcal{D}}_{\text{decoder}}\left(\sum_{m=1}^{512} \beta_m \tau_m\right) \approx \underbrace{E'_{\text{ext}}(z' \equiv z'_o)}_{\text{prediction}}, \quad (4)$$

where $\mathcal{D}$ denotes the decoder mapping from the latent response to the predicted field profile. Unless otherwise specified, rectified linear units (ReLU) are used in all convolutional and fully connected layers. Further details are listed in the Table 1 in Appendix A.

## *2.3 Data structure*

The training dataset follows the workflow established in our previous study [11], but is now extended across multiple function families to broaden coverage and stress-test cross-family generalization. For clarity, we briefly recap the EFISH forward formulation and note several assumptions that guide data generation.

We adopt the EFISH forward formulation (equation (1)) written as:

$$P^{(2\omega)} \propto [\alpha^{(3)} \cdot N \cdot P_o^{(\omega)}]^2 \cdot E_o^2 \cdot z_R \cdot \left| \int_{-\infty}^{\infty} \frac{E'_{\text{ext}}(z') \cdot e^{iuz'}}{1 + i \cdot z'} \, dz' \right|^2, \quad (5)$$

where

$$E'_{\text{ext}}(z') = \frac{E_{\text{ext}}(z')}{E_o}, \quad z' = \frac{z}{z_R}, \quad u = \Delta k \cdot z_R.$$

Here, the electric field $E'_{\text{ext}}(z')$ is normalized by the peak amplitude $E_o$, and thus encodes only the shape of the profile. Other quantities (e.g., $z_R$, $\Delta k$, and the prefactors in brackets) are treated as constants between calibration and measurement, known *a priori*, or readily quantified [6,11]. In practice, a calibration curve (obtained at a single point along the beam path) suffices to recover $E_o$; thereafter, $E'_{\text{ext}}$ can be rescaled to obtain the absolute field. For simplicity, distinct polarization effects of the signal are neglected here, although they are essential for resolving distinct field vector components (typically $E_x$ and $E_y$) whose magnitudes and spatial signatures may differ substantially. As these components can exhibit different underlying shapes (and amplitudes), training a single model to predict both with high fidelity would require additional effort. Accordingly, we continue our focus on the vertical (dominant) component of the electric field, $E_y$, for which we deem the present model to be most relevant. An additional constraint is that our model has been trained exclusively on symmetric profiles and is thus expected to be applicable only to symmetric field distributions. In this regard, we propose a symmetry index ($\mathcal{SI}$) (discussed in Appendix C) which can assess the suitability of our model for a given reconstruction task. Ultimately, we anticipate that constraints of symmetry do not limit its capability, and with the appropriate training, can recover any arbitrary field shape.

To improve the training dataset quality, we expand upon our previous Fuzzy family of $E'_{\text{ext}}(z')$ field shapes (both single- and double-peak) [6,11], to include the Voigt function as shown in Table 2 of Appendix B. Motivated by its applications in spectroscopy and use in EFISH studies [18,19,33], the Voigt function represents a convolution of a Lorentzian and Gaussian profile, two *independent* families with markedly different shapes [34]. This extended diversity of the training dataset encompasses a wider variety of profile shapes, enabling the model to achieve better generalizability and performance on EFISH inversion tasks.

Across both families, the non-dimensional parameters $\{a, b, c\}$ (normalized by $z_R$) govern physically meaningful attributes and are varied within prescribed ranges [11]. Specifically, $a$ controls the half-width at half-maximum (HWHM), $b$ adjusts the flank steepness, and $c$ sets the dip depth or peak separation. The Voigt profile is implemented via a standard pseudo-Voigt approximation, i.e., a linear combination of a Gaussian and a Lorentzian component [34,35]; the Gaussian width $\sigma$ and Lorentzian half-width $\gamma$ used in sampling are listed in Appendix B. A nonzero $c$ yields double-peak profiles akin to the low-field core observed behind ionization wave fronts [6]. Following our prior studies, we sample $a \in [2,10]$, $b \in [1,3]$, and $c \in [0,5]$ (Appendix B), a design shown to span a broad continuum of physically relevant shapes [11].

Subsequently, each $E'_{\text{ext}}(z')$ is inserted into the (forward EFISH) equation (5) with an additional axial displacement ($z'_o$) representing the translation between the applied electric field with respect to the probe beam focus:

$$P^{(2\omega)}(z_o') \propto [\alpha^{(3)} \cdot N \cdot P_o^{(\omega)}(z_o')]^2 \cdot E_o^2 \cdot z_R \cdot \left| \int_{-\infty}^{\infty} \frac{E'_{\mathrm{ext}}(z'-z_o') \cdot e^{iuz'}}{1+i \cdot z'} \, \mathrm{d}z' \right|^2, \qquad (6)$$

with $z_o' = \frac{z_o}{z_R}$. Following Yang et al. [11], the displacement range is bounded by practical constraints to avoid interference between the discharge geometry and optical components (e.g., lens obstruction), and we set $z_o' \in [-50, 50]$ (The range and values of $z'$ for describing $E'_{\mathrm{ext}}(z')$ are identical to $z_o'$ and thus for convenience, $z' \equiv z_o'$). Because the EFISH response depends on profile shape rather than absolute amplitude, all signals are peak-normalized and denoted as $P_{\mathrm{norm}}^{2\omega}$ herein; in evaluating equation (6), only the squared magnitude of the integral is needed. For numerical stability and consistency, $z_o'$ is max-normalized to $[-1, 1]$ and $u$ is normalized by its maximum value (i.e. $u' = \frac{u}{u_{\max}}$). In total, we generate 703,682 profile pairs $\left(E'_{\mathrm{ext}}(z') \leftrightarrow P_{\mathrm{norm}}^{(2\omega)}(z_o')\right)$ across all families, each sampled at 109 points, yielding $\sim 7.7 \times 10^7$ data points for training and evaluation.

Data augmentation is pivotal for accuracy and robustness [36,37]. We employ two complementary strategies. First, we crop 20% of randomly selected input profiles, $P_{\mathrm{norm}}^{2\omega}$: the cropping window spans 50% of the profile length with a randomized start index. Cropped profiles $\{[P_{\mathrm{norm}}^{2\omega}, z_o'], [u']\}$ are then upsampled via linear interpolation to match the input resolution and reincorporated into the training set, exposing the model to 'partial-view' scenarios common in practice (e.g., limited data or incomplete profiles). Second, we inject controlled additive Gaussian noise into 20% of inputs in each batch using a jitter layer [38,39]: specifically, zero-mean noise with standard deviation STD = 5% of the input amplitude. This mimics sensor noise and background fluctuations frequently observed in EFISH measurements. Together, these augmentations consistently improve performance in our parametric studies, boosting accuracy and cross-family generalization while enhancing resilience to partial (or incomplete) profiles and noise.

### *2.4 Training details*

The DDON model takes $P_{\mathrm{norm}}^{2\omega}$ as an input and predicts the electric field shape $E'_{\mathrm{ext}}(z')$. The 703,682-profile dataset is randomly split into three groups: 80% training, 10% validation, and 10% test. The validation set supports hyperparameter tuning and checkpoint selection, while the held-out test set provides an unbiased estimate of generalization.

Training was performed with an Adam (initial learning rate $1 \times 10^{-3}$) and a reduce-on-plateau scheduler that halves the learning rate after 5 consecutive epochs without improvement on the validation metric. We used a batch size of 512 and optimized the mean squared error (MSE) loss. Early stopping terminated training if the validation loss failed to improve for 15 successive epochs, with a maximum of 100 epochs overall [11,40]. In practice, this training regimen yielded stable optimization, striking a balance between accuracy and the regularization needed for reliable inversion. Models were implemented in Keras with a TensorFlow backend and trained on the NUS Vandar high-performance computing cluster.

### *2.5 Integrated Gradients (IG)*

Over the past decade, a variety of attribution methods have been proposed to explain predictions from complex models by assigning importance scores to input features. Recent work has clarified shared principles (e.g., completeness, sensitivity) that unify many of these techniques. Among gradient-based approaches, IG stands out for its inherent interpretability and robustness [26]. To elucidate the decision-making of our DDON, we adopt IG to quantify how each element of the EFISH input (signal) influences the predicted electric field.

Formally, IG quantifies the contribution of each input feature by integrating the gradient of the model output with respect to the input along a straight path from a baseline to the actual input. In this study, we use a zero baseline ($\left\{P_{\mathrm{norm,baseline}}^{(2\omega)}\right\} = 0$)—an array of zeros matching the input EFISH shape $\left\{P_{\mathrm{norm}}^{(2\omega)}\right\}$–which corresponds to a null electric field and serves as a neutral, featureless reference. For each EFISH profile $P_{\mathrm{norm}}^{2\omega}(z_o')$, we construct interpolants along the path:

$$P^{(\alpha)} = P_{\mathrm{norm,baseline}}^{(2\omega)} + \alpha \times \left[P_{\mathrm{norm}}^{(2\omega)} - P_{\mathrm{norm,baseline}}^{(2\omega)}\right], \qquad \alpha \in [0,1], \qquad (7)$$

where $\alpha = 0$ corresponds to the baseline and $\alpha = 1$ the actual input. Letting $F$ (i.e. $\mathcal{D}(\mathcal{G})$) denote the learned mapping from input to the predicted $E'_{\mathrm{ext}}(z')$, the attribution for feature of $P_{\mathrm{norm}}^{(2\omega)}(z'_{o,i})$ at each location $z'_{o,i}$ ($i = 1 \ldots 109$) can be integrated along the path (equation 7) as:

$$IG_{z'_{o,i}} = \left(P_{z'_{o,i}} - P_{z'_{o,i}}^{\mathrm{baseline}}\right) \int_{\alpha=0}^{\alpha=1} \frac{\partial F(P^{(\alpha)})}{\partial P_{z'_{o,i}}} \, \mathrm{d}\alpha, \qquad (8)$$

with $IG_{z'_{o,i}}$ the integrated gradient for the $i$th input location of $P_{\mathrm{norm}}^{(2\omega)}$. Accordingly, a higher $IG_{z'_{o,i}}$ index denotes a higher attribution of EFISH at this location to the final E-field predictions, and vice versa. We approximate the path integral with 100 uniform Riemann steps ($\alpha = 0{:}\,0.01{:}\,1$), balancing computational cost and accuracy. The resulting IG maps are overlaid on the original inputs, highlighting the regions that most strongly influence the model's output, enabling a systematic interpretation of the learned representations (e.g., see Figure 4a).

## **3. Results and Discussion**

### *3.1 Generalizability*

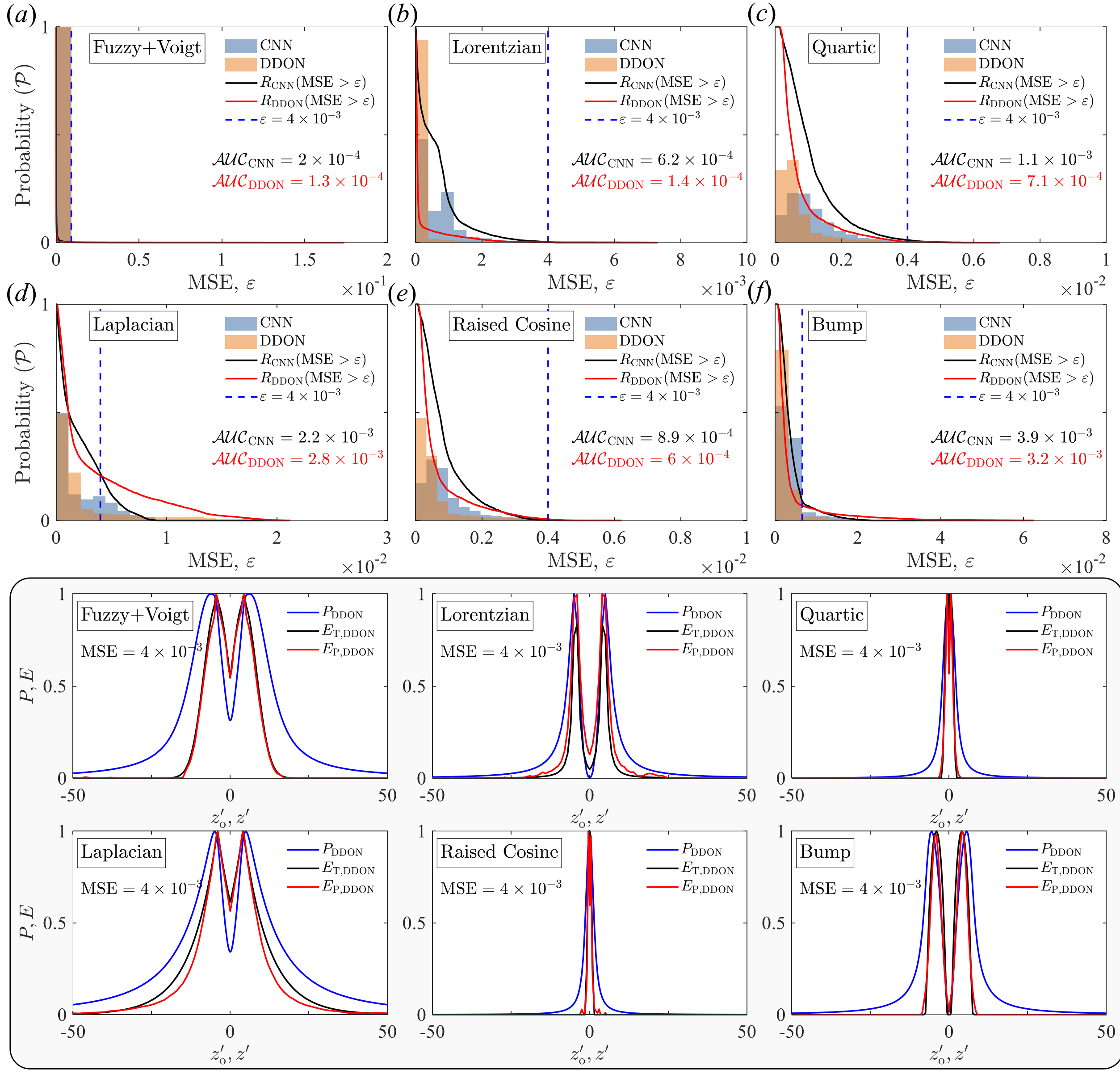

Figure 2 Probability distribution of the performance and sample prediction results for DDON and CNN [11] across six different function spaces: (a) Fuzzy+Voigt, (b) Lorentzian, (c) Quartic, (d) Laplacian, (e) Raised Cosine, and (f) Bump. The normalized EFISH signal (model input) is henceforth denoted by $P$ , while $E_{\mathrm{T}}$ and $E_{\mathrm{P}}$ henceforth indicate the normalized true and model-predicted electric field profiles respectively (DDON vs. CNN). Red and black ROC-like curves summarize comparative performance over the test ensembles. For visual clarity, six representative cases with $\varepsilon(\mathrm{MSE}) = 4 \times 10^{-3}$ are shown.

### *3.1.1 Function-space generalizability.*

To assess the function-space-generalizability of the DDON model (henceforth abbreviated as DDON), we evaluate it on unseen profiles from six function spaces representing $E'_{\mathrm{ext}}(z')$—(i) Fuzzy + Voigt, (ii) Lorentzian, (iii) Quartic, (iv) Laplacian, (v) Raised Cosine, and (vi) Bump—and compare its performance against our previous CNN model (henceforth abbreviated as CNN) used in [11]. It is important to emphasize that of these six, four of them ((ii) to (vi)) correspond to function spaces which the model has not been trained on. In general, these functions span a large spectrum of common bell-shaped (and double-peak) profiles, thus providing a rigorous measure of model generalizability across different function families.

As shown in Figure 2(a–g), DDON outperforms the CNN across almost all six function spaces (each space contains 9,191 profiles), yielding markedly lower prediction errors. To ensure a fair comparison, it should be noted that the CNN model is retrained on the same exact dataset as the DDON. This manifests as left-shifted mean-squared-error (MSE) distributions for DDON, with a higher concentration of low-error predictions than the CNN. For a quantitative comparison, we plot a Receiver-operating characteristic (ROC-like tail-probability curve), $R(\varepsilon) = \mathcal{P}(\mathrm{MSE} > \varepsilon)$, i.e., the probability that a prediction's error exceeds a threshold $\varepsilon$. A lower $R(\varepsilon)$ indicates that a smaller fraction of predictions exceed the tested error threshold, while a larger $R(\varepsilon)$ indicates that a majority of the predictions' errors exceed the threshold. We also report the area under this ROC curve, $\mathcal{AUC} = \int_0^{\varepsilon_{\max}} R(\varepsilon)\, \mathrm{d}\varepsilon$, computed over a fixed $[0, \varepsilon_{\max}]$ for both models; smaller $\mathcal{AUC}$ values signify better accuracy across thresholds. Consistent with the empirical MSE histograms, DDON exhibits faster-decaying $R(\varepsilon)$ and smaller $\mathcal{AUC}$ than the CNN in most spaces (especially its performance at predicting small MSE); see Figure 2(a–g). Taken together, these results indicate that the DDON confers stronger robustness when extrapolating to previously unseen profile families, particularly those with sharp features or heavy tails.

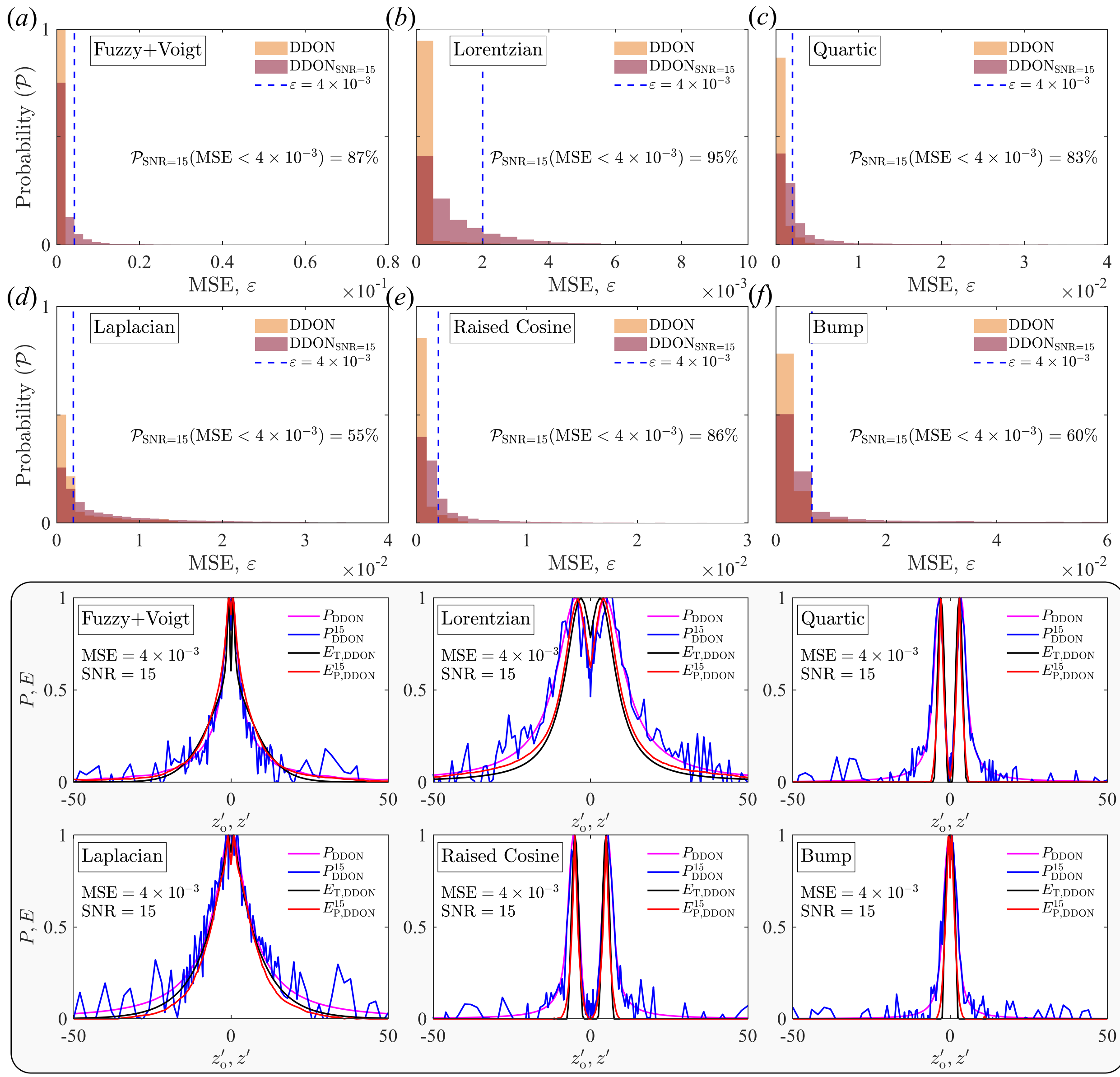

Figure 3 Robustness of DDON to additive Gaussian noise across six function spaces (following the same format as Figure 2): (a) Fuzzy+Voigt, (b) Lorentzian, (c) Quartic, (d) Laplacian, (e) Raised Cosine, (f) Bump. Signal-to-noise ratio fixed at SNR = 15. Superscripts in $P^{15}$ (noise-corrupted EFISH input) and $E_{\mathrm{T}}^{15}$ (DDON prediction) indicate evaluation at SNR = 15.

To visualize profile-specific performance, representative predictions (via DDON) with $\mathrm{MSE} \approx 4 \times 10^{-3}$ (arbitrarily

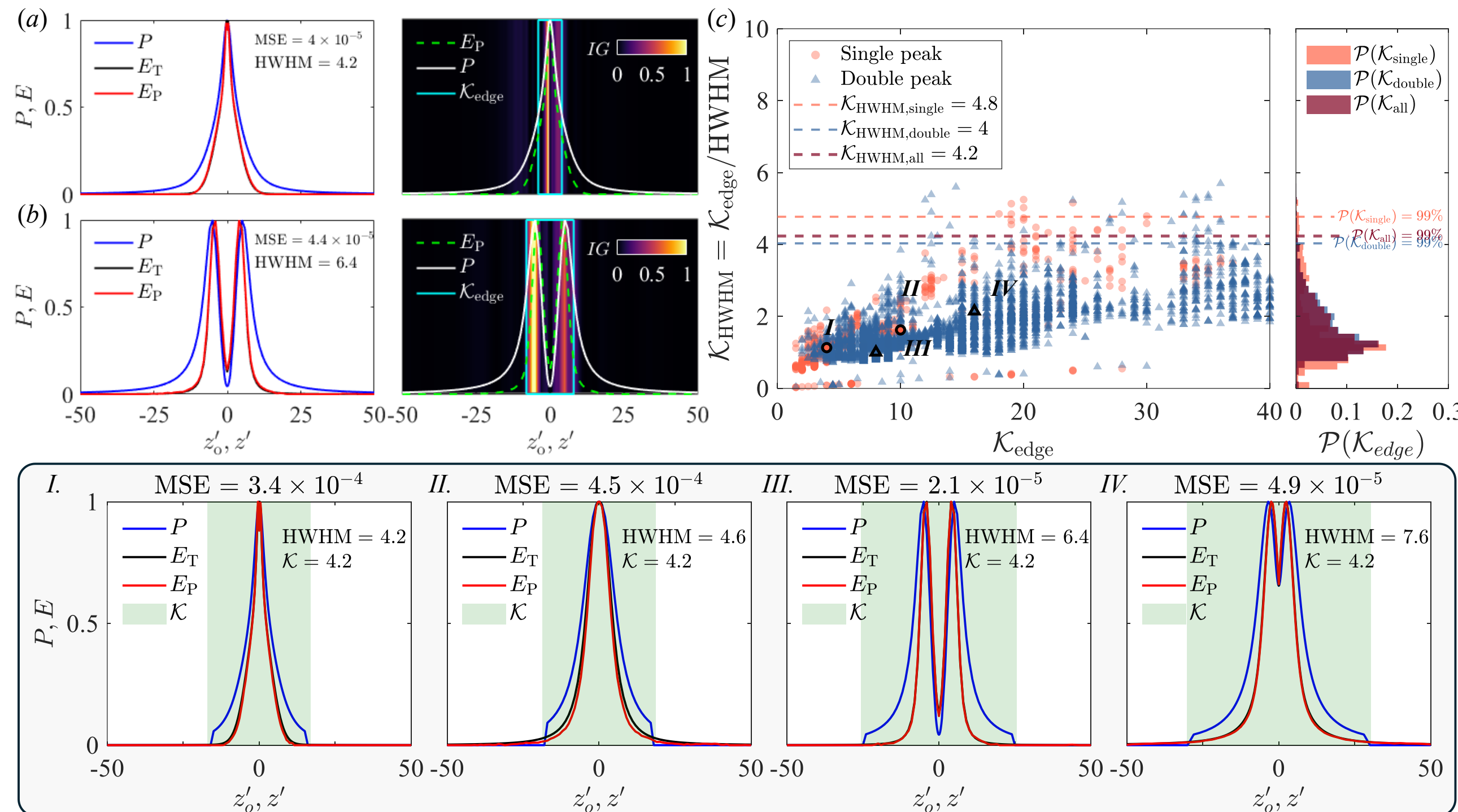

Figure 4 (a,b) Integrated-gradient (IG) attribution heatmap for DDON (single-peak and double-peak profiles), highlighting input regions most influential on the prediction. (c) Identification of a universal, normalized EFISH sampling window ($\mathcal{K}_{\mathrm{HWHM,all}} = 4.2$) from the distribution of 'key-feature boundary locations ($\mathcal{K}_{\mathrm{edge}}$) linking EFISH inputs to DDON outputs. (I-IV) Sample E-field predictions using sparsified inputs restricted to this IG-identified, key-feature window (denoted in the plot legend as $\mathcal{K}$ and shaded in green here, as well the plots that follow).

selected) are also displayed in Figure 2. These examples are largely representative of the overall accuracy between the two models. Using $\varepsilon = 4 \times 10^{-3}$ as a threshold, the possibility of predictions satisfying $\mathrm{MSE} \leq \varepsilon$ are 99.7%, 99.8%, 99.6%, 79.2%, 99.3%, and 86.1% for the Fuzzy + Voigt, Lorentzian, Quartic, Laplacian, Raised Cosine, and Bump function spaces, respectively.

- ***Feasibility of a mathematical reconstruction method.***

We attempt to benchmark DDON against a classical reconstruction method based on Fourier deconvolution, following Park et al. [41]. This method is capable of successfully reconstructing local plasma emission intensity from a chord-integrated emission profile, analogous to our path-integrated EFISH signal. In brief, the chord-integrated signal is transformed to frequency space (through an FFT), divided by an optical geometry kernel, and then deconvolved to recover the local intensity distribution. While effective for linear integration signals, this strategy appears ill-suited to EFISH inversion (equation 6). The EFISH forward mapping comprises a coherence integral followed by a modulus operation, which introduces strong nonlinearity and erases phase information that is essential for recovering the underlying electric field. This mismatch underscores the intrinsic difficulty of EFISH inversion and motivates research into novel approaches (such as DDON in this case) for a good solution.

### *3.1.2 Noise robustness.*

Beyond function-space generalization, we evaluate the DDON's robustness to noise—an unavoidable aspect of real experimental measurements. Our goal is to assess whether the model maintains predictive accuracy and stability when inputs are corrupted by measurement noise. To this end, we inject controlled Gaussian white noise into the EFISH input profiles, yielding a noisy set denoted $P_{\mathrm{DDON}}^{15}$ with $\mathrm{SNR} = 15$ (corresponding to a Peak signal-to-noise ratio PSNR $\approx$ 18 dB), and compare its performance against the clean-input baseline $P_{\mathrm{DDON}}$. This evaluation probes the model's ability to generalize under noisy conditions, highlighting its reliability in realistic settings.

As shown in Figure 3, DDON maintains strong performance even under substantial noise perturbations. For most function families, the probability that a prediction satisfies $\mathrm{MSE} \leq 4 \times 10^{-3}$ remains high: $\mathcal{P}(\mathrm{MSE} \leq 4 \times 10^{-3}) \gtrsim 85\%$ for $P_{\mathrm{DDON}}^{15}$, with the exceptions of the Laplacian and Bump families, where the probability is relatively lower $\mathcal{P}(\mathrm{MSE} \leq 4 \times 10^{-3}) \gtrsim 55\%$. These results indicate that DDON is inherently noise-robust for a broad range of profiles, although improvements in performance for the Laplacian and Bump families could further enhance model robustness. The same robustness test ($\mathrm{SNR} = 15$) is also applied to the CNN, where it is observed to perform poorly (not shown) under this large noise level, with the possibility of predictions satisfying $\mathrm{MSE} \leq 4 \times 10^{-3}$ to be 64%, 25%, 30%, 41%, 30%, 9.6% for the six function spaces, respectively.

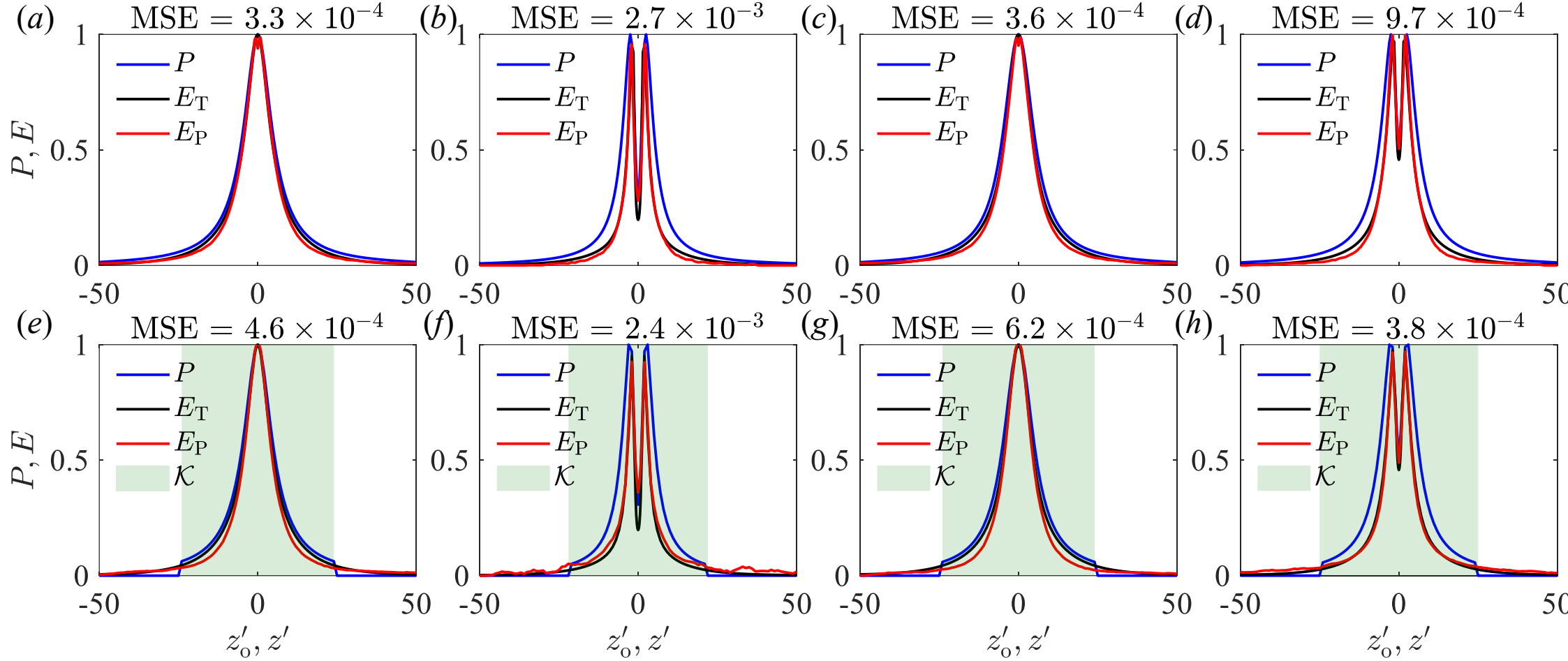

Figure 5 Simulated ($E_{\mathrm{T}}$-reproduced from Ref. [6]) versus DDON-predicted ($E_{\mathrm{P}}$) electric-field profiles in a pin-plane electrode geometry for an applied voltage of 50 kV. The discharge E-field is probed 5 mm (a, b, e, f) and 8 mm (c, d, g, h) below the pin tip at varying time instants. Panels (a-d) use the complete EFISH input profile, whereas panels (e-h) use an 'artificially sparsified' EFISH profile with (valid) points restricted to the key-feature window.

Furthermore, one acknowledges that noise can arise in other forms beyond Gaussian noise, such as from optical components, contamination from plasma emission, and shot noise. Therefore, we further expand our noise robustness tests to cover these scenarios by employing DDON on the same function spaces across Perlin noise ($\mathrm{SNR} = 15$), and Poisson noise ($\mathrm{SNR} = 15$). The former (Perlin noise) generates spatially correlated, continuous patterns that mimic the "cloud-like" artifacts characteristic of plasma emissions and optics components. The latter captures the signal-dependent nature of shot noise. In both tests (results shown in Appendix D), the DDON continues exhibiting well generalization across function spaces to different noise sources with minor degradation in performance compared to the Gaussian noise case. This suggests that its learned features are fundamental to the underlying signal structure rather than being overfitted to a specific Gaussian noise source profile. Perhaps the ultimate proof of the above this is evidenced by the model's exemplary performance on experimental data as described in sections 3.3.2 and 3.3.3 that follow.

### *3.2 Model interpretability*

As aforementioned, IG provides a systematic way to visualize portions of the input which most strongly influences a model's output. Here, we apply IG to 5,000 randomly selected test cases to examine how EFISH inputs affect the predicted E-field across distinct function families.

As illustrated in Figure 4, the IG heat maps highlight the input regions that drive the predictions. For single-peak profiles (Figure 4a), the highest attributions cluster near the beam focus (i.e., the profile center) and taper toward the tails—suggesting, in practice, denser sampling near the beam focus and sparser coverage outside. For double-peak profiles (Figure 4b), the central contribution diminishes and attribution shifts to the off-center maxima, indicating that informative structure resides primarily around the flanking peaks. These trends echo that of Ref. [16], who reported central dominance for Gaussian-like fields, motivating the search for a 'key-feature" region that contains the dominant information.

To delineate this 'key-feature' region, we compute the gradient magnitude of each IG heatmap along the axial $z$-coordinate and locate the first (i.e., farthest from the beam focus) prominent jump using Otsu's method [42]. This threshold-detection method systematically examines the variance between successfully separated regions under every possible threshold value, and the (threshold) value yielding the maximum between-region variance is assigned as the optimal threshold for boundary detection. We denote the $z$-location of this detected boundary as a 'key-feature' boundary, or $\mathcal{K}_{\mathrm{edge}}$ (cyan lines in Figure 4). To quantify a universal sampling window (common to all profiles), we extract for *every profile*, the location of its 'key-feature' boundary, $\mathcal{K}_{\mathrm{edge}}$, and normalize this value by the input EFISH half-width at half-maximum (HWHM), mathematically defined as $\mathcal{K}_{\mathrm{HWHM}} = \frac{\mathcal{K}_{\mathrm{edge}}}{\mathrm{HWHM}}$. Remarkably, these values all cluster under $\mathcal{K}_{\mathrm{HWHM}} \lesssim 5$, implying that sampling within $\pm 5$ HWHM of the focus for any profile is generally sufficient for the DDON to produce reliable E-field predictions (Figure 4c). More specifically, examining the 99th percentile for each class, we find that sampling within $\mathcal{K}_{\mathrm{HWHM}} = 4.8$ for single-peak profiles and $\mathcal{K}_{\mathrm{HWHM}} = 4.0$ for double-peak profiles suffices for accurate prediction with DDON. Therefore, as a practical guideline, researchers may consider targeting $\mathcal{K}_{\mathrm{HWHM}} = 4.2$ when planning measurements or when an expected HWHM of the EFISH signal is available. In a typical EFISH experiment, the exact boundaries of such a data acquisition window can be identified based on the location of the peak signal.

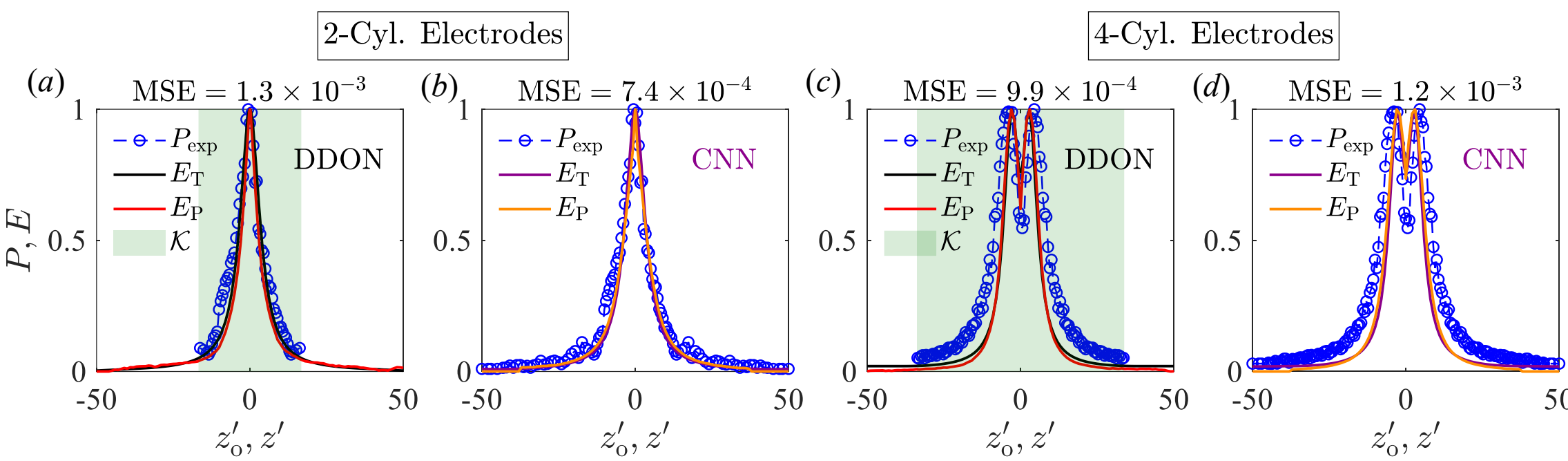

Figure 6 Predicted electric-field profiles ($E_{\mathrm{P}}$) for two separate electrode configurations sustained under electrostatic conditions based on Ref. [11]. (a,b) Single-peak case: one electrode pair; (c,d) double-peak case: two electrode pairs separated axially (along the $z$-axis). Circle markers (blue) indicate EFISH sampled points within the IG-informed window (green). (a,c) are DDON predictions while (b,d) are CNN predictions reprodued from Ref. [11].

We validate this guideline by selecting four representative cases with different HWHM values (Figure 4I-IV) and performing an 'ablation study'. The latter is a test commonly used in the ML literature to assess the contribution(s), or sensitivity, of a specific input feature(s) to the final output/result. In our context, an ablation test involves assigning a null value to the EFISH signals lying outside the key region identified by IG (i.e. $\mathcal{K}$) and feeding this modified profile into DDON. The output profile, $E_{\mathrm{P}}(z')$, is then compared against the corresponding ground truth profile, $E_{\mathrm{T}}(z')$. This provides a direct indication as to whether EFISH signals acquired beyond this region substantially affect the predicted/underlying field profile.

To this end, the model is tested on sparsified EFISH profiles that satisfy a chosen general criterion $\mathcal{K}_{\mathrm{HWHM,all}} = 4.2$. As shown in Figure 4(I-IV), DDON retains strong predictive performance for both single- and double-peak profiles, closely tracking the ground truth. This confirms that the IG-identified window captures the dominant information needed for accurate reconstruction and effectively emulates missing-data scenarios. In the following section, we corroborate these findings with additional simulations and experiments, emphasizing their utility for planning efficient, reliable measurements.

The above result is not entirely unexpected. In the reverse instance, one notes that different regions of a given field profile, $E_{\mathrm{ext}}(z)$, are weighted differently in terms of their influence on $P^{(2\omega)}(z)$. And it is thus conceivable, that $z$-locations away from the field peak (for instance, for a single-peak profile) could have a negligible effect on the resulting EFISH signal.

As a cautionary note, we remark that the IG-method is a universal XAI technique that may, in principle, be applied to many AI algorithms, particularly datasets in which derivatives exist. However, the quantitative values of the key feature regions, such as $\mathcal{K}_{\mathrm{HWHM}} = 4.2$, are likely to be model-specific (i.e., unique to DDON), rather than universally applicable to any EFISH experiment.

### *3.3 Application to Simulation and Experimental Datasets*

In this section, we apply DDON to several realistic plasma systems to reconstruct several unknown E-field distributions from data comprising both simulations and experiments.

#### *3.3.1 Discharge simulations.*

A convenient aspect of using numerical data is that the electric field information is directly available and can be assumed as the ground truth. To this end, we apply DDON to previously published numerical data obtained via a well-established, *axisymmetric* fluid model [43]. Briefly, a positive polarity discharge is initiated in a canonical pin-plane geometry via a pulsed waveform consisting of a rise time of 0.5 ns and applied voltage plateau of 50 kV. The time evolution of the electric field profile is tracked at vertical offsets of 5 mm and 8 mm below the (high-voltage) pin anode. Following Ref. [6], these numerically-predicted profiles are taken to be the ground truth (i.e., $E_{\mathrm{T}}$), and used to generate their corresponding EFISH profiles, $P$, via equation (6). These generated profiles are in turn fed into DDON to yield a separate set of predicted electric field profiles, $E_{\mathrm{P}}$. A comparison is then made between $E_{\mathrm{T}}$ and $E_{\mathrm{P}}$ as shown in Figure 5. Using complete EFISH profiles (full spatial coverage), DDON reconstructions closely track the simulated fields (i.e., $E_{\mathrm{T}}$) reported by Chng et al. [6]; see Figure 5(a–d). Notably, when we apply the IG-informed sampling window with $\mathcal{K}_{\mathrm{HWHM,double}} = 4.0$ ($\mathcal{K}_{\mathrm{HWHM,all}} = 4.2$ for cases e,g) and zero the remaining regions, DDON retains its predictive performance: the MSE between reconstructions and their ground truth remains essentially unchanged relative to the full-profile case, consistent with the sampling guidance established in § 3.2.

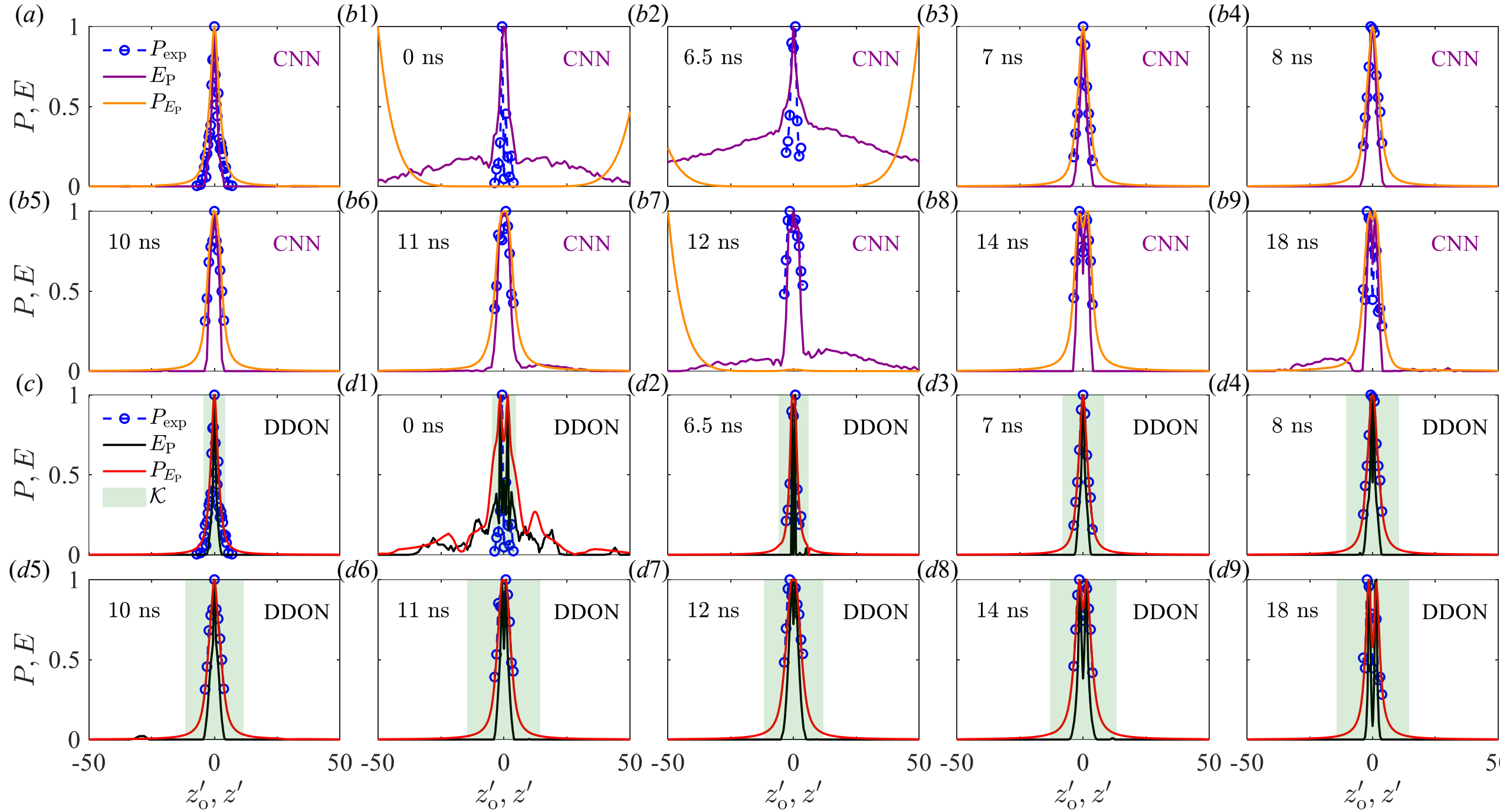

Figure 7 Predicted electric-field profiles via CNN ($E_{\mathrm{P}}$, purple) and DDON ($E_{\mathrm{P}}$, black) for a pin-pin electrode geometry at different applied voltages: (a,c) 3.5 kV – no breakdown (electrostatic field); (b,d) 8.2 kV – corona discharge, shown across nine temporal snapshots along with their respective forward transforms, $P_{E_{\mathrm{P}}}$. Circle markers (blue) indicate experimental EFISH data while the green shaded regions denote the IG-informed window.

### *3.3.2 Experimental electrostatic fields.*

To evaluate DDON on purely electrostatic conditions, we apply this new model to measurements from our previous experiments [11]. The single-peak profile was generated using a single pair of cylindrical electrodes, while the double-peak profile was sustained by two electrode pairs separated along the $z$-axis. Additional details of the setup can be found in Appendix B of Ref. [11].

Figure 6 shows reconstructions of $E'_{\mathrm{ext}}(z')$obtained from data sampled within the IG-informed key-feature window (i.e., $\mathcal{K}_{\mathrm{HWHM,all}} \approx 4.2$ for the single-peak case in Figure 6a and $\mathcal{K}_{\mathrm{HWHM,double}} \approx 4.0$ for the double-peak case in Figure 6c). Ground-truth fields, $E_{\mathrm{T}}$, are computed via 2D Laplacian simulations for each electrode configuration. DDON predictions closely track the simulated fields in both cases, indicating strong applicability to real electrostatic data. The prediction accuracy (i.e. MSE) shows a similar level to that of our previous work [11], as indicated in Figure 6(b,d). Here, the marginally better performance of the CNN model could be due to the fact that these E-field profiles better correspond to the training dataset used in that model; this tradeoff between accuracy and generalizability is typical of most ML approaches. Another observation is that, even if the measured EFISH input exhibits slight asymmetry, the current (symmetry-trained) DDON tends to produce symmetric reconstructions. This is especially helpful for the prediction of symmetric field distributions because the presence of noise may often bias a true symmetric signal towards an asymmetric one under certain conditions. However, this also highlights a present limitation and motivates further study of asymmetric profiles.

### *3.3.3 Nanosecond pulsed corona discharge.*

To assess the model's performance on an experimental plasma, we apply DDON to EFISH measurements in a pin-pin electrode geometry under both Laplacian conditions, and in a discharge at ambient pressure. More details of this setup can be found in [44,45]. Briefly, a pair of tungsten pin electrodes (tip diameters < 160 um) are separated by a 10 mm vertical interelectrode gap and driven by 10 ns (FWHM) pulses at a repetition rate of 10 kHz. An applied voltage of 3.5 kV sustains a Laplacian field while a higher amplitude of 8.2 kV generates a pulsed corona discharge. The electric field is probed at a location of about 600 μm directly below the high voltage (HV) electrode. (The electrodes are aligned along the vertical $x$-axis, and EFISH measurements are conducted in the centre of the horizontal $y$ - $z$ plane along the $z$ -axis.) To facilitate the field reconstruction, EFISH signal profiles are acquired by sequentially translating the electrode system relative to the laser focus in steps of $\Delta z = 0.5 - 1$ mm ($\Delta z'_{\mathrm{o}} \approx 1$). Other pertinent parameters of this experiment are identical to that in [11]: Rayleigh range, $z_{\mathrm{R}} = 1.35$ mm; the wave-vector mismatch, $\Delta k = 0.5\ \mathrm{cm}^{-1}$ gives $u = -\Delta k \cdot z_{\mathrm{R}} \approx -0.068$.

Table 1 Configurations of the Decoder-DeepONet (DDON). The general structures are as depicted in Figure 1. AE is short for autoencoder. For Residual block, the stride is 2 with zero-padding employed simultaneously.

| **Subnet** | **Layer** | **Size** | **Num layers** |
|---|---|---|---|
| Branch net | Input layer | $[*,109,1]$ | 1 |
| | Jitter | $[*,109,1]$ | 1 |
| | Residual block | $[*,109]\times 64$<br>$[*,55]\times 128$<br>$[*,28]\times 256$ | 3 |
| | Average pooling | $[*,14]\times 256$ | 1 |
| | Flattening | $[*,3584]\times 1$ | 1 |
| | Dense layer (AE) | $[*,\{1024,128,32,128,512\}]\times 1$ | 5 |
| Trunk net | Input layer | $[*,109,2]$ | 1 |
| | Residual block | $[*,109]\times 64$<br>$[*,55]\times 128$<br>$[*,28]\times 256$ | 3 |
| | Average pooling | $[*,14]\times 256$ | 1 |
| | Flattening | $[*,3584]\times 1$ | 1 |
| | Dense layer | $[*,512]\times 1$ | 3 |
| Latent space | $\text{Dense}_{(\text{Brunch}\times\text{Trunk})}$ | $[*,512]\times 1$ | 1 |
| Decoder | Dense layer | $[*,512]\times 1$ | 2 |
| | | $[*,256]\times 1$ | 2 |
| | | $[*,109]\times 1$ | 2 |
| **Hyperparameter setting** | | **Value** | |
| Learning rate (decaying) | | $1\times 10^{-3}$ (initial) | |
| Batch size | | 512 | |
| Epochs | | 100 (early stopping) | |

Figure 7 summarizes the reconstruction results (denoted as $E_{\mathrm{P}}$), and provides a comparison between the predictions of DDON and CNN. Unlike the previous sections, the true field, $E_{\mathrm{T}}$, is unknown for these experiments. Nonetheless, we can still perform an independent accuracy check by forward-propagating $E_{\mathrm{P}}$ through the EFISH integral (equation 6) to produce $P_{E_{\mathrm{P}}}$ (red curve) and compare it with the measured signal $P_{\exp}$ (blue curve). *This represents an important advantage of this approach – the accuracy of an ML prediction can still be evaluated even if the underlying 'true' profile is unknown.* Similar to the previous section, we benchmark DDON ($E_{\mathrm{P}}$, black curve; $P_{E_{\mathrm{P}}}$, red curve) against CNN ($E_{\mathrm{P}}$, purple; $P_{E_{\mathrm{P}}}$, orange) which accepts a full range EFISH profile requiring, for certain datasets, extrapolation to the boundaries (not shown) as an input. In several cases (i.e., time instants), the predicted field profiles for CNN lead to significant deviations between $P_{E_{\mathrm{P}}}$ and $P_{\exp}$. Conversely, this agreement (between $P_{E_{\mathrm{P}}}$ and $P_{\exp}$) is excellent under both electrostatic and discharge conditions for DDON, despite severe sampling sparsity (as few as eight EFISH spatial points). This close match between the measured and 'forward-predicted' EFISH signals provides strong validation of the reconstructed fields and demonstrates DDON's robustness even under demanding experimental conditions. We add that, for consistency, regions of $z'_{\mathrm{o}}$ in which experimental data is unavailable is assigned a value of zero, without any need for extrapolation as in the case of CNN. Notably, all acquisition points fall within the IG-informed key-feature window (e.g., $\mathcal{K}_{\mathrm{HWHM,all}} \approx 4.2$; see § 3.2), consistent with our sampling guideline and likely contributing to the observed fidelity. We stress that this window is deliberately conservative: accurate reconstructions can sometimes be obtained without sampling out to the boundary, as IG-identified features do not necessarily lie exactly at the edge in Figure 4c. These results highlight such flexibility while reinforcing the utility of the IG-based criterion as a practical guide for experimental planning. A current limitation is that model performance may degrade for highly irregular or asymmetric inputs (e.g., Figure 7b1). Even then, DDON typically returns physically plausible profiles rather than catastrophic failures, which further attests to its robustness.

Table 2 Summary of the function families used for training (i.e., Fuzzy and Voigt) and quantifying the generalizability (i.e., all families) of DDON and CNN (from Ref.[11]).

| **Function family** | **Mathematical expression** | **Parameter** |
|---|---|---|
| Fuzzy | $E(z;a,b,c)=\frac{1}{1+\lvert\frac{\lvert z\rvert-c}{a}\rvert^{b}}$ | $a\in(0{,}10]$, $b\in[1{,}3]$, $c\in[0{,}5]$ |
| Voigt | $\begin{cases} G(z;\sigma,b,c)=\frac{1}{\sigma\sqrt{2\pi}}\exp\left(-\frac{(z-c)^{b}}{2\sigma^{2}}\right); \\ L(z;\gamma,b,c)=\frac{1}{\pi}\frac{\gamma}{(z-c)^{b}+\gamma^{2}}; \\ E(z;\sigma,\gamma,b,c)=\int_{-\infty}^{\infty}G\,(z';\sigma,b,c)\,L(z-z';\gamma,b,c)\,\mathrm{d}z' \end{cases}$ | $\sigma\in(0{,}10]$, $\gamma\in(0{,}10]$, $b\in[1{,}3]$, $c\in[0{,}5]$ |
| Lorentzian | $E(z;a,c)=\frac{1}{\pi}\frac{a}{(z-c)^{2}+a^{2}}$ | $a\in(0{,}10]$, $c\in[0{,}5]$ |
| Quartic | $E(z;a,c)=\begin{cases}\frac{15}{16\,a}\left(1-\left(\frac{z-c}{a}\right)^{2}\right)^{2}, & \lvert z-c\rvert\le a \\ 0, & \lvert z-c\rvert>a\end{cases}$ | $a\in(0{,}10]$, $c\in[0{,}5]$ |
| Laplacian | $E(z;a,c)=\frac{1}{2a}\exp\left(-\frac{\lvert z-c\rvert}{a}\right)$ | $a\in(0{,}10]$, $c\in[0{,}5]$ |
| Raised Cosine | $E(z;a,c)=\begin{cases}\frac{1}{2a}\left[1+\cos\left(\pi\frac{z-c}{a}\right)\right], & \lvert z-c\rvert\le a \\ 0, & \lvert z-c\rvert>a\end{cases}$ | $a\in(0{,}10]$, $c\in[0{,}5]$ |
| Bump | $E(z;a,c)=\begin{cases}\exp\left(-\frac{a^{2}}{a^{2}-(z-c)^{2}}\right), & \lvert z-c\rvert<a \\ 0, & \lvert z-c\rvert\ge a\end{cases}$ | $a\in(0{,}10]$, $c\in[0{,}5]$ |

While not explicitly discussed in detail, we add that standard post-processing of the raw experimental EFISH data (as with most EFISH experiments) prior to applying DDON is helpful for improving the reconstruction accuracy. For the experiments discussed above (and in § 3.3.2), subtraction of background plasma emission and SHG from optical components as well as signal averaging over multiple laser shots, was performed before using these as inputs to the model.

### *3.3.4 Additional limitations and the issue of solution uniqueness*

The preceding sections (in particular § 2.3) have addressed the various limitations associated *directly* with the prediction of the reconstructed electric field profile using DDON. In this penultimate section, we highlight a few other limitations of this model, apart from those earlier discussed, which *indirectly* impact the solution accuracy.

DDON addresses the fundamental transform (or inverse problem) between $P^{(2\omega)}(z_{\mathrm{o}}')$ and $E_{\mathrm{ext}}'(z')$ fundamentally described in equation (6). In this context, variables such as the gas hyperpolarizability ( $\alpha^{(3)}$ ), neutral density ( $N$ ), and wavevector mismatch ($\Delta k$) have been assumed constant when constructing our model. This assumption may not always be valid when considering differences between calibration and a plasma [6]. More importantly, while **uniform** changes in these other parameters are likely to be adequately accommodated by DDON, accounting for such **non-uniform** variations along the $z$-axis (i.e., direction of laser propagation) will demand further model development.

Finally, while we have emphasized the forward transform (via equation (6)) as a vital way of assessing the accuracy of a reconstructed profile, this utility is evidently predicated on the need for uniqueness between $P^{(2\omega)}(z_{\mathrm{o}}')$ and $E_{\mathrm{ext}}'(z')$ – a common feature of many inverse problems. We are currently unable to make a definitive statement regarding the uniqueness between these two variables, which would thus imply that these reconstructions are in effect, 'most probable' solutions. We should clarify that this notion of probability is not because valid DDON reconstructions do not satisfy equation (6), but rather, that more than one solution profile to equation (6) *could* exist.

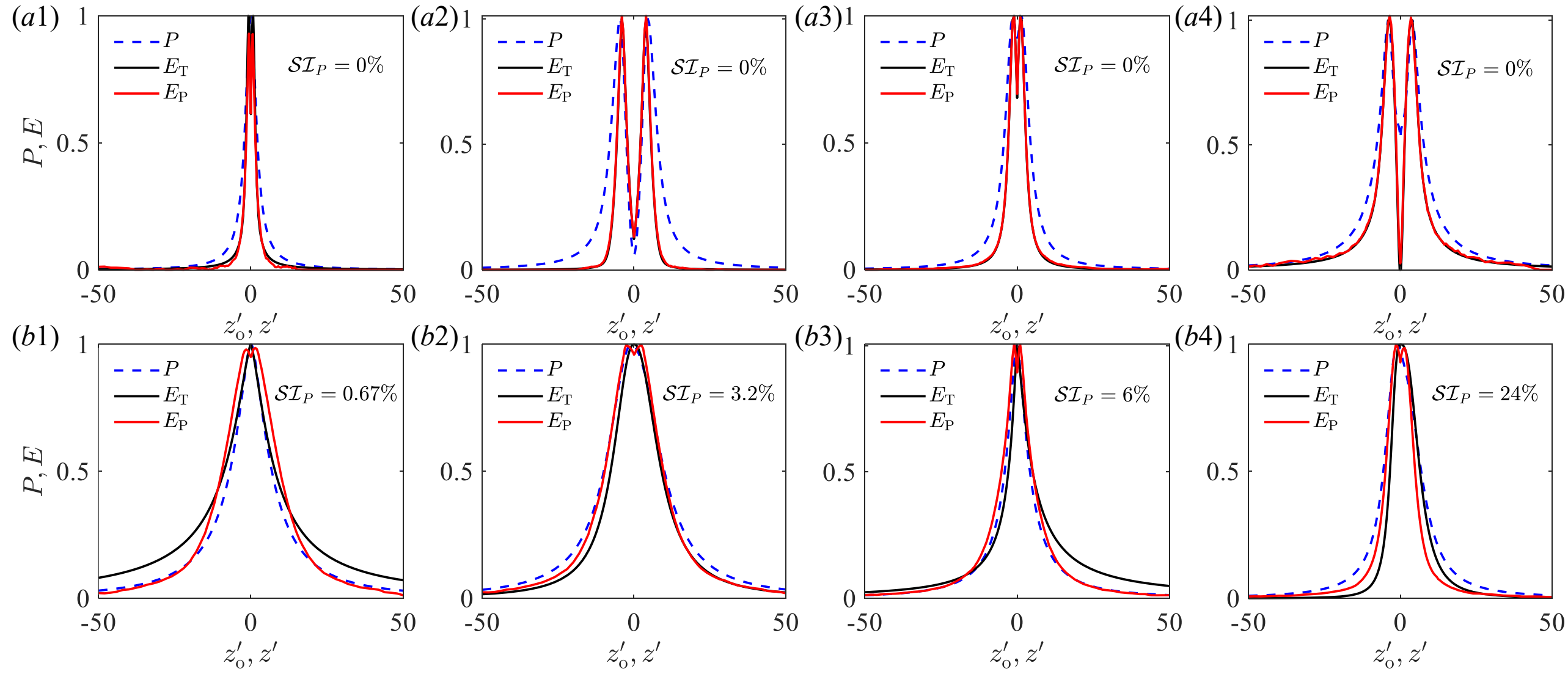

Figure 8 Symmetry analysis for EFISH profiles (as applicable for electric field), including (a) symmetry profiles and (b) asymmetry profiles. $\mathcal{SI} > 0$ denotes asymmetry profiles for a noise-free distribution.

Nonetheless, we believe the chances of this non-uniqueness (i.e., multiple solutions) to be rather low for the following reasons – (i) phase-compensating differences (due to the modulus in equation (6)) that result in multiple solutions are likelier to exist at a single point, but less so for an entire profile; and (ii) ongoing work suggests that additional constraints of profile symmetry and bell-shaped topology (i.e., $E'_{\text{ext}}(z') \to 0$ for $|z'| \to \infty$) tend to suppress the possibility of non-unique solutions. We anticipate further research on this topic of uniqueness for assessing the effects of experimental error, variations in the (far) tails of the field distribution and the non-linear $z$-weighting of the Gouy phase shift.

## 4 Conclusion

In this work, we introduce a novel Decoder-DeepONet (DDON) operator-learning framework for reconstructing electric-field profiles from EFISH measurements. Following an existing approach, this reconstruction method acquires an EFISH signal profile along the axis of laser propagation, $P^{(2\omega)}(z)$, and feeds this into our ML model to obtain the shape of the unknown electric field, $E_{\text{ext}}(z)$.

Unlike our earlier CNN, DDON couples ResNet encoders with a compact autoencoder and an operator head, enabling it to learn *function-to-function* mappings (not possible with CNN) – in this case the complex inverse relationship, $P^{(2\omega)}(z) \to E_{\text{ext}}(z)$ in equation (6). It is this feature that allows DDON to maintain high predictive accuracy even when applied to unseen function families of $E_{\text{ext}}(z)$, beyond those that the model has originally been trained on.

We have compared DDON's performance against our earlier CNN trained on the same dataset. In almost all instances, DDON consistently exhibits superior out-of-distribution generalization, maintaining high reconstruction accuracy on unseen profiles from six distinct function families (four of these six are unseen) used to represent $E_{\text{ext}}(z)$, as well as when subjected to high noise levels (SNR $= 15$, PSNR $\approx 18$ dB) typical of an EFISH experiment. Most encouragingly, DDON continues to perform well when applied to simulation and experimental datasets. In particular, the model successfully recovers the underlying field profile shape from EFISH measurements in a nanosecond pulsed corona discharge. It is worth noting that this reconstruction is achieved even in EFISH profiles with as few as 8 datapoints. This is especially useful given that classical mathematical approaches can be extremely sensitive to sparse datasets. It is well-known that SNR issues often affect the quality of data near the tail(s) of a profile for instance, and this may compromise the accuracy of such methods.

Another distinctive contribution of this work is interpretable and actionable guidance for data acquisition: using the method of Integrated Gradients (IG), we identify input regions that most strongly affect reconstruction accuracy. Crucially, this allows us to quantify a normalized sampling window, $\mathcal{K}_{\text{HWHM}} = 4.2$, which based on our DDON, provides a high probability of reconstruction success. This conservative estimate is systematically validated on existing datasets and confirmed to good effect, demonstrating its reliability in practice.

Nonetheless, several limitations of this model warrant mention. Notably, it is only applicable to bell-shaped profiles with an existing axis of symmetry, a drawback we believe can be addressed with appropriate training. This will in turn extend the model's applicability beyond the dominant component of the electric field, and permit reconstruction of antisymmetric profiles associated with the often-weaker orthogonal component. Finally, the need to sample multiple points is likely to rule out the possibility of obtaining instantaneous, single-shot data and will likely limit its use to reproducible or quasi-steady discharges.

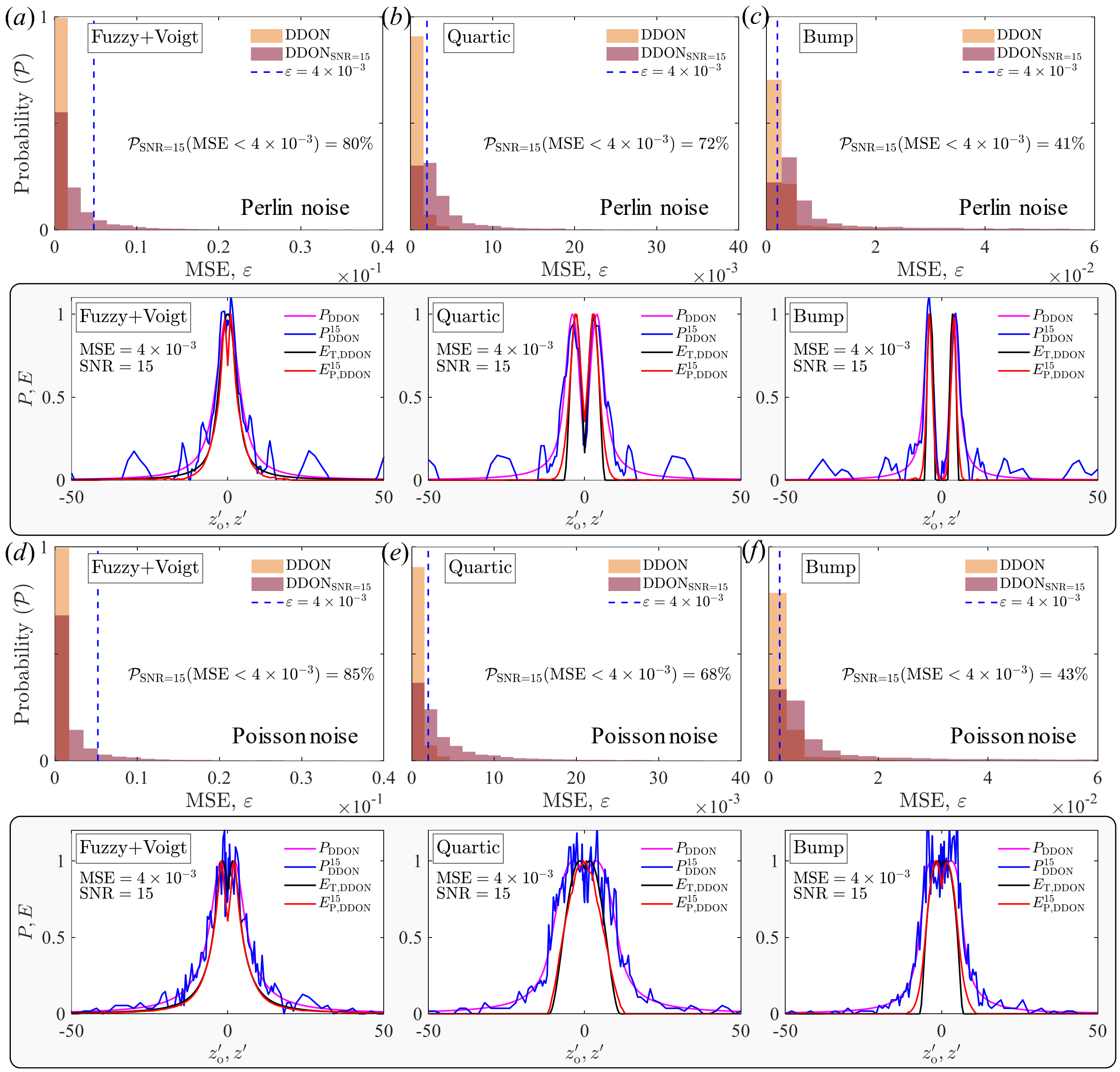

Figure 9 Robustness of DDON to additive Perlin noise (a-c) and Poisson noise (d-f) across three function spaces (following the same format as Figure 3): (a,d) Fuzzy+Voigt, (b,e) Quartic, (e,f) Bump. Global signal-to-noise ratio is fixed at SNR = 15.

In closing, we reiterate an important point: as mentioned in § 3.3.3, any predicted field profile from DDON can be immediately mapped back through equation (6) to obtain a corresponding EFISH signal, $P(z)$, assuming a unique relationship between these two variables. By comparing this predicted signal directly against the experimentally acquired profile, the model's accuracy can be instantly verified *before* the reconstructed field profile is used for further analysis. This crucial check makes the deployment of DDON risk-free and gives us further confidence in its applicability to experimental plasma studies.

## Acknowledgements

We wish to acknowledge the support from a Singapore Ministry of Education (MOE) Academic Research Fund Tier 1 grant (22-5447-A0001), and an MOE PhD research scholarship for Mr. Edwin Setiadi Sugeng. We are also grateful for useful suggestions from Prof. Uwe Czarnetski of Ruhr University Bochum, Prof. Sanghoo Park of Korea Advanced Institute of Science and Technology, as well as Prof. Gianmarco Mengaldo and Mr. Qi-Yang Er of the National University of Singapore.

## Appendix A. Hyperparameters of DDON

In this appendix, we list details of the hyperparameters of the DDON with best performance (see Table 1). All hyperparameters are selected after several rounds of parametric optimization.

## Appendix B. Function families

This appendix describes the function families used for training, validation, test, and qualification in this study (see Table 2), including: (i) Fuzzy (same as that in Ref. [11]), (ii) Voigt, (iii) Lorentzian, (iv) Quartic, (v) Laplacian, (vi) Raised Cosine, and (vii) Bump. These families are expected to cover a wide variety of existing bell-shaped and double-peak profiles. As mentioned in the main text, parameter $a$ determines the characteristic width of the profile (HWHM) and $b$ affects the profile steepness. Parameter $c$ controls the distance between the two peaks for double-peak profiles. Increasing $c$ enlarges the depth and width of the dip between two peaks.

## Appendix C. Symmetry screening of EFISH signals

To ascertain if an EFISH profile (or the underlying electric field) is effectively symmetric, we suggest the symmetry index ($\mathcal{SI}$) [46–48], a robust metric for discretely sampled data:

$$\mathcal{SI} = \frac{1}{N}\sum_{i=1}^{N}\left|\frac{z_{\mathrm{r},i} - z_{\mathrm{l},i}}{1/2\,\left(z_{\mathrm{r},i} + z_{\mathrm{l},i}\right)}\right| \times 100\%,$$

where $N$ is the number of symmetric pairs, and $z_{\mathrm{r},i}$, $z_{\mathrm{l},i}$ are matched sampling points to the right and left of the central reference (e.g., EFISH peak or optical focus). As a strict rule: $\mathcal{SI} = 0\%$ denotes high symmetry, and $\mathcal{SI} > 0\%$ denotes asymmetry (Figure 8). For noise-free signals, $\mathcal{SI}$ reliably separates symmetric from asymmetric profiles; under substantial noise, however, $\mathcal{SI}$ can be biased upward and yield false positives. While a noise-robust, symmetry criterion lies beyond the scope of this paper, denoising, median-based pairing, and uncertainty-aware thresholds (e.g., bootstrap confidence intervals) are promising directions.

With these caveats, we propose a straightforward screening procedure to identify profiles suitable for DDON inversion: (i) verify symmetry of the electrode system geometry; (ii) obtain a reference, 'noise-less' electric field distribution via a Laplacian simulation for this geometry and synthesize the EFISH signal using equation (6); (iii) compute $\mathcal{SI}$ for the target EFISH profile or electric field; (iv) if $\mathcal{SI}$ indicates effective symmetry (e.g., $\mathcal{SI} = 0\%$), proceed with DDON-based reconstruction, see Figure 8. In our tests, DDON remains reliable even for experimentally acquired signals at $\mathrm{SNR} = 15$, provided the underlying profile is expected to be symmetric (via the above screening procedure), offering a practical path to signal inversion in real-world measurements.

## Appendix D. Robustness of DDON to other noise sources

In this appendix, we show results (see Figure 9) of our robustness test of DDON to two other different noise sources – Perlin and Poisson noise - in addition to Gaussian noise, as an attempt to mimic experimental noise from plasma emission and optical components, as well as signal-dependent shot noise, respectively.

## References

[1] Bigio I J, Finn R S and Ward J F 1975 Electric-Field Induced Harmonic Generation as a Probe of the Focal Region of a Laser Beam *Appl. Opt., AO* **14** 336–42

[2] Dogariu A, Goldberg B M, O'Byrne S and Miles R B 2017 Species-Independent Femtosecond Localized Electric Field Measurement *Phys. Rev. Appl.* **7** 024024

[3] Goldberg B M, Chng T L, Dogariu A and Miles R B 2018 Electric field measurements in a near atmospheric pressure nanosecond pulse discharge with picosecond electric field induced second harmonic generation *Applied Physics Letters* **112** 064102

[4] Chng T L, Starikovskaia S M and Schanne-Klein M-C 2020 Electric field measurements in plasmas: how focusing strongly distorts the E-FISH signal *Plasma Sources Sci. Technol.* **29** 125002

[5] Goldberg B M, Hoder T and Brandenburg R 2022 Electric field determination in transient plasmas: in situ & non-invasive methods *Plasma Sources Sci. Technol.* **31** 073001

[6] Chng T L, Pai D Z, Guaitella O, Starikovskaia S M and Bourdon A 2022 Effect of the electric field profile on the accuracy of E-FISH measurements in ionization waves *Plasma Sources Sci. Technol.* **31** 015010

[7] Raskar S, Orr K, Adamovich I V, Chng T L and Starikovskaia S M 2022 Spatially enhanced electric field induced second harmonic (SEEFISH) generation for measurements of electric field distributions in high-pressure plasmas *Plasma Sources Sci. Technol.* **31** 085002

[8] Vorenkamp M, Steinmetz S A, Chen T Y, Ju Y and Kliewer C J 2023 Suppression of coherent interference to electric-field-induced second-harmonic (E-FISH) signals for the measurement of electric field in mesoscale confined geometries *Opt. Lett., OL* **48** 1930–3

[9] Hogue J, Cusson P, Meunier M, Seletskiy D V and Reuter S 2023 Sensitive detection of electric field-induced second harmonic signals *Opt. Lett., OL* **48** 4601–4

[10] Billeau J-B, Cusson P, Dogariu A, Morozov A, Seletskiy D V and Reuter S 2024 Coherent homodyne detection for amplified cross-beam electric-field induced second harmonic *Appl. Opt.* **63** 5203

[11] Yang Z, Sugeng E S and Chng T L 2025 A deep learning approach for electric field profile reconstruction based on the E-FISH method *Plasma Sources Sci. Technol.*

[12] Lepikhin N D, Luggenhölscher D and Czarnetzki U 2025 Increasing the EFISH sensitivity by an optical parametric amplifier *Opt. Lett., OL* **50** 5809–12

[13] Nakamura S, Sato M, Fujii T and Kumada A 2022 Measurement Method for Electric Field in Streamer Discharge Based on Electric-Field-Induced Second-Harmonic Generation *2022 IEEE International Conference on Plasma Science (ICOPS)* 2022 IEEE International Conference on Plasma Science (ICOPS) pp 1–1

[14] Guo Y 2025 Measurement of the electric field distribution in streamer discharges *Phys. Rev. Res.* **7**

[15] Chen S, He H, Chen Y, Liu Z, Xie S, Che J, He K and Chen W 2024 Measurement of inhomogeneous electric field based on electric field-induced second-harmonic generation *Measurement* **231** 114576

[16] Luo Q, Chen X, Zhang Y, Xie J and Lan L 2025 A discrete inversion measurement method for non-uniform electric fields based on electric field induced second harmonic generation *Optics & Laser Technology* **192** 113939

[17] Sato T, Umemoto T, Sato M, Fujii T and Kumada A 2025 Phase-resolved measurement of electric-field-induced second harmonics and its application to noninvasive electric field sensing *Plasma Sources Sci. Technol.* **34** 105019

[18] van der Schans M, Böhm P, Teunissen J, Nijdam S, IJzerman W and Czarnetzki U 2017 Electric field measurements on plasma bullets in N2 using four-wave mixing *Plasma Sources Sci. Technol.* **26** 115006

[19] Huang B, Zhang C, Zhu W, Lu X and Shao T 2021 Ionization waves in nanosecond pulsed atmospheric pressure plasma jets in argon *High Voltage* **6** 665–73

[20] Lu L, Jin P and Karniadakis G E 2021 DeepONet: Learning nonlinear operators for identifying differential equations based on the universal approximation theorem of operators *Nat Mach Intell* **3** 218–29

[21] Lu L, Jin P, Pang G, Zhang Z and Karniadakis G E 2021 Learning nonlinear operators via DeepONet based on the universal approximation theorem of operators *Nat Mach Intell* **3** 218–29

[22] He J, Koric S, Kushwaha S, Park J, Abueidda D and Jasiuk I 2023 Novel DeepONet architecture to predict stresses in elastoplastic structures with variable complex geometries and loads *Computer Methods in Applied Mechanics and Engineering* **415** 116277

[23] Chen B, Wang C, Li W and Fu H 2024 A hybrid Decoder-DeepONet operator regression framework for unaligned observation data *Physics of Fluids* **36** 027132

[24] Ahn S, Bae J, Yoo S and Nam S K 2025 Deep transfer operator learning for predicting low temperature plasma sheath dynamics in semiconductor processing *Phys. Plasmas* **32** 093505

[25] Ali S, Abuhmed T, El-Sappagh S, Muhammad K, Alonso-Moral J M, Confalonieri R, Guidotti R, Del Ser J, Díaz-Rodríguez N and Herrera F 2023 Explainable Artificial Intelligence (XAI): What we know and what is left to attain Trustworthy Artificial Intelligence *Information Fusion* **99** 101805

[26] Sundararajan M, Taly A and Yan Q 2017 Axiomatic Attribution for Deep Networks *Proceedings of the 34th International Conference on Machine Learning* International Conference on Machine Learning (PMLR) pp 3319–28

[27] Mei Y, Zhang Y, Zhu X, Gou R and Gao J 2024 Fully Convolutional Network-Enhanced DeepONet-Based Surrogate of Predicting the Travel-Time Fields *IEEE Transactions on Geoscience and Remote Sensing* **62** 1–12

[28] He K, Zhang X, Ren S and Sun J 2016 Deep Residual Learning for Image Recognition *2016 IEEE Conference on Computer Vision and Pattern Recognition (CVPR)* 2016 IEEE Conference on Computer Vision and Pattern Recognition (CVPR) (Las Vegas, NV, USA: IEEE) pp 770–8

[29] LeCun Y, Bengio Y and Hinton G 2015 Deep learning *Nature* **521** 436–44

[30] Gu J, Wang Z, Kuen J, Ma L, Shahroudy A, Shuai B, Liu T, Wang X, Wang G, Cai J and Chen T 2018 Recent advances in convolutional neural networks *Pattern Recognition* **77** 354–77

[31] Kiranyaz S, Avci O, Abdeljaber O, Ince T, Gabbouj M and Inman D J 2021 1D convolutional neural networks and applications: A survey *Mechanical Systems and Signal Processing* **151** 107398

[32] Srivastava N, Hinton G, Krizhevsky A, Sutskever I and Salakhutdinov R 2014 Dropout: a simple way to prevent neural networks from overfitting *J. Mach. Learn. Res.* **15** 1929–58

[33] Sato T, Sogame M, Sato M, Fujii T, Oishi Y and Kumada A 2025 Remote measurement of electric fields based on electric-field-induced second-harmonic generation using a nanosecond laser at a distance of 10 meters *Optics Express* **33** 37192–203

[34] Ida T, Ando M and Toraya H 2000 Extended pseudo-Voigt function for approximating the Voigt profile *J Appl Cryst* **33** 1311–6

[35] Thompson P, Cox D E and Hastings J B 1987 Rietveld refinement of Debye–Scherrer synchrotron X-ray data from Al2O3 *J Appl Cryst* **20** 79–83

[36] Bonzanini A D, Shao K, Graves D B, Hamaguchi S and Mesbah A 2023 Foundations of machine learning for low-temperature plasmas: methods and case studies *Plasma Sources Sci. Technol.* **32** 024003

[37] Taylor L and Nitschke G 2018 Improving Deep Learning with Generic Data Augmentation *2018 IEEE Symposium Series on Computational Intelligence (SSCI)* 2018 IEEE Symposium Series on Computational Intelligence (SSCI) pp 1542–7

[38] Holmstrom L and Koistinen P 1992 Using additive noise in back-propagation training *IEEE Trans. Neural Netw.* **3** 24–38

[39] An G 1996 The Effects of Adding Noise During Backpropagation Training on a Generalization Performance *Neural Comput* **8** 643–74

[40] Bengio Y 2012 Practical recommendations for gradient-based training of deep architectures

[41] Park H, Choe W and Yoo S J 2010 Spatially resolved emission using a geometry-dependent system function and its application to excitation temperature profile measurement *Spectrochimica Acta Part B: Atomic Spectroscopy* **65** 1029–32

[42] Otsu N 1975 A Tlreshold Selection Method from Gray-Level Histograms *Automatica* **11** 23–7

[43] Bourdon A, Péchereau F, Tholin F and Bonaventura Z 2020 Study of the electric field in a diffuse nanosecond positive ionization wave generated in a pin-to-plane geometry in atmospheric pressure air *J. Phys. D: Appl. Phys.* **54** 075204

[44] M Alkhalifa A, Di Sabatino F, A Steinmetz S, Pfaff S, Huang E, H Frank J, J Kliewer C and A Lacoste D 2024 Quantifying the thermal effect and methyl radical production in nanosecond repetitively pulsed glow discharges applied to a methane-air flame *J. Phys. D: Appl. Phys.* **57** 385204

[45] Alicherif M, Sugeng E, Yang Z, Lacoste D A and Chng T L 2026 Toward Quantitative Electric-Field Measurements of Inception Clouds in Nanosecond Discharges Using E-FISH Assisted by Machine Learning

[46] Robinson R O, Herzog W and Nigg B M 1987 Use of force platform variables to quantify the effects of chiropractic manipulation on gait symmetry *J Manipulative Physiol Ther* **10** 172–6

[47] Nigg S, Vienneau J, Maurer C and Nigg B M 2013 Development of a symmetry index using discrete variables *Gait & Posture* **38** 115–9

[48] Bermudez D H and Culberson W 2025 Correlative Symmetric Index: An alternative mathematical evaluation for beam profile symmetry